\documentclass[aps,superscriptaddress,10pt,nofootinbib,notitlepage,twocolumn,prd]{revtex4-1}

\usepackage{hyperref}
\usepackage{xcolor}
\usepackage{graphicx}
\usepackage{physics}
\usepackage{bm}
\usepackage{amssymb}

\newcommand{\be}[1]{\begin{equation} #1 \end{equation}}

\newcommand{\SM}[1]{\text{\sc sm}}
\newcommand{\KKLT}[1]{\text{$\mathrm{KKLT}$}}
\newcommand{\LVS}[1]{\text{$\mathrm{LVS}$}}

\def\be{\begin{equation}}
\def\ee{\end{equation}}

\begin{document}

\title{
 The spectrum of $n_s$ constraints from DESI and CMB data
}

\author{Evan McDonough}
\affiliation{Department of Physics, University of Winnipeg, Winnipeg MB, R3B 2E9, Canada}

\author{Elisa G. M. Ferreira}
\affiliation{Kavli IPMU (WPI), UTIAS, The University of Tokyo, 5-1-5 Kashiwanoha, Kashiwa, Chiba 277-8583, Japan}
\affiliation{Center for Data-Driven Discovery, Kavli IPMU (WPI), UTIAS, The University of Tokyo, Kashiwa, Chiba 277-8583, Japan}

\begin{abstract}
We present the spectrum of $n_s$ constraints from current CMB data (Planck, ACT, SPT-3G) combined with DESI BAO data, and highlight the interplay of $n_s$ with the optical depth to reionization $\tau$.
The spectral index $n_s$ of the primordial power spectrum provides a window into early universe, and constraints on $n_s$ play an important role in discriminating early universe models such as models of cosmic inflation. Historically constrained by cosmic microwave background (CMB) experiments, the constraints on $n_s$ shift upward when CMB data is combined with the latest baryon acoustic oscillation (BAO) data from the Dark Energy Spectroscopic Instrument (DESI). 
Recent work explained the origin of this and the relation to the BAO-CMB tension between CMB experiments and DESI BAO, and as a case study presented constraints on $n_s$ from the combination of Atacama Cosmology Telescope (ACT) DR6 data and DESI DR2 data.
Here we present constraints from Planck (PR3 and PR4), ACT, the South Pole Telescope (SPT), and the combination of all three CMB experiments, CMB-SPA, with and without DESI DR2 BAO data, and with and without CMB lensing data.
In all cases the constraint on $n_s$ is shifted upwards when DESI is included, with the largest shift exhibited by ACT. This is accompanied by a commensurate shift in the constraint on the optical depth to reionization $\tau$, which is again greatest for ACT.
When CMB data are combined into CMB-SPA and combined with DESI 
the $n_s$ constraint disfavors at more than $2\sigma$ the inflation models preferred by Planck alone, such as Higgs, Starobinsky, and exponential $\alpha$-attractors, in favor of other models, such as polynomial $\alpha$-attractors.
This work motivates the further study of the tension between CMB and DESI BAO data, and of the rich interplay between $n_s$ and $\tau$.
\end{abstract}

\maketitle

\section{Introduction}
\label{sec:intro}
\parskip 5pt

The spectral index $n_s$ connects the $\Lambda$CDM cosmological model to the primordial universe. Accordingly, $n_s$ is a powerful discriminator of models of cosmic inflation. This is evidenced by a flurry of model building activity (c.f. \cite{Kallosh:2025rni,Maity:2025czp,Heidarian:2025drk,Frolovsky:2025iao,Chakraborty:2025jof,Lizarraga:2025aiw,Cacciapaglia:2025xqd,Addazi:2025qra,Kim:2025dyi,Pallis:2025epn,Odintsov:2025wai,Gialamas:2025kef,McDonald:2025odl,Yin:2025rrs,Aoki:2025wld,Antoniadis:2025pfa,Haque:2025uis,He:2025bli,Salvio:2025izr,Pallis:2025nrv,Haque:2025uga,Hai:2025wvs,Saini:2025jlc,Kohri:2025lau,Liu:2025qca,Drees:2025ngb,Zharov:2025evb,Okada:2025lpl,Byrnes:2025kit,Yogesh:2025wak,Peng:2025bws,McDonald:2025tfp,Choudhury:2025vso,Mohammadi:2025gbu,Gialamas:2025ofz,Lynker:2025wyc,Linde:2025pvj,Aoki:2025ywt,Ketov:2025cqg,Kallosh:2025ijd,Odintsov:2025eiv,Zahoor:2025nuq,Oikonomou:2025xms,Oikonomou:2025htz,Modak:2025bjv,German:2025ide,Karananas:2025fas,Yuennan:2025tyx,Hell:2025lbl,Allegrini:2025jha,Iacconi:2025odq,Ghoshal:2025ejg,Yuennan:2025mlg,Ellis:2025zrf,Pal:2025xbt,Huang:2025xyf,Kallosh:2025sji,Qiu:2025uot}) following the result from the Atacama Cosmology Telescope \cite{ACT:2025fju}  and South Pole Telescope \cite{SPT-3G:2025bzu} that the latest CMB data combined with the latest Baryon Acoustic Oscillation (BAO) data from the Dark Energy Spectroscopic Survey \cite{DESIDR2} disfavor at greater than $2\sigma$ the inflation models that were taken to be the benchmark models of future CMB experiments such as CMB-S4 and LiteBIRD \cite{Chang:2022tzj}, namely the Starobinsky model, Higgs inflation, and exponential $\alpha$-attractor models, in favor of other models.

Ref.~\cite{Ferreira:2025lrd} explained the shifting landscape of $n_s$ constraints in terms of the tension between CMB and DESI BAO data sets, termed the BAO-CMB tension. The CMB experiments exhibit varying but mild tension with DESI \cite{SPT-3G:2025bzu}: 2.0$\sigma$ for Planck, 3.1$\sigma$ for ACT, $2.5\sigma$ for SPT,  3.7$\sigma$ for SPT+ACT, and 2.8$\sigma$ for SPT+Planck+ACT (``CMB-SPA''). Ref.~\cite{Ferreira:2025lrd} demonstrated that $n_s$ is correlated with the BAO parameters in the fit of $\Lambda$CDM to CMB data. This leads to a shift in $n_s$ when CMB and DESI data are combined despite the BAO-CMB tension.

In this work, we consider the full suite of latest CMB experiments including Planck, ACT, SPT, and their combination, CMB-SPA. We present constraints on $n_s$ and the BAO parameters, $r_d h$ and $\Omega_m$, with and without DESI DR2 BAO, and with and without CMB lensing. We present the first constraints from SPT primary CMB combined with DESI, the first constraints from SPT, CMB lensing, and DESI, using the multifrequency SPT likelihood, and the first constraints from CMB-SPA, with and without DESI, using the multifrequency (not \texttt{lite}) ACT likelihood.

We find that all CMB experiments exhibit a correlation between $n_s$ and BAO parameters $r_d h$ and $\Omega_m$, with a strength that tracks the relative sensitivity to large angular scales. All CMB experiments, alone and in the combination CMB-SPA, exhibit a shift in $n_s$ when combined with DESI. The largest shift in posterior mean is exhibited by ACT, which can be anticipated, since ACT has the largest tension with DESI \cite{SPT-3G:2025bzu}. On the other hand, Planck has the smallest uncertainty on $n_s$.  When ACT and SPT are combined with Planck into CMB-SPA and combined with DESI, this gives a constraint on $n_s$ that disfavors the Higgs, Starobinsky, and exponential $\alpha$-attractor inflation models at greater than $2\sigma$, consistent with \cite{SPT-3G:2025bzu}.

These results indicate that the BAO-CMB tension is the driving force behind shifts in $n_s$ when CMB and DESI BAO data are combined.
This motivates further investigation of the source of the BAO-CMB  tension, and further experimental efforts to improve the sensitivity to $n_s$. 

We note that while CMB B-modes are often thought to be the ``Holy Grail'' \cite{Abazajian:2013vfg} of inflation\footnote{See \cite{Brandenberger:2011eq} for an alternative view.}), many well motivated inflation models predict a tensor-to-scalar ratio that falls far below the threshold for next generation CMB experiments (see e.g.~\cite{Lorenzoni:2024krn,McDonough:2020gmn,Wolf:2024lbf}). These models would be more directly tested by precision measurements of $n_s$ and by the running $\alpha_s$ \cite{Lorenzoni:2024krn,Hardwick:2018zry,Martin:2024nlo}. Thus it may well be that $n_s$, and $\alpha_s$, continue to lead the way in testing models of cosmic inflation.

\section{Data sets and methodology}
\label{sec:datasets}

Our analysis will make use of the following CMB data sets:
\begin{itemize}
     \item Atacama Cosmology Telescope (ACT): ACT DR6 primary CMB \footnote{\url{https://act.princeton.edu/act-dr6-data-products}} \cite{ACT:2025fju} and lensing \cite{ACT:2023kun,ACT:2023dou} data. Following the ACT convention, when analysing ACT we include the Planck low-$\ell$ EE \texttt{sroll2} likelihood.
     \item South Pole Telescope (SPT): SPT-3G D1, \footnote{\url{https://pole.uchicago.edu/public/Home.html}, \url{https://github.com/SouthPoleTelescope/spt_candl_data}} \cite{SPT-3G:2025bzu, muse, Balkenhol:2024sbv}.  Following the SPT convention (see \S VII C and Appendix I in \cite{SPT-3G:2025bzu} for details) when analysing SPT data we impose a Gaussian prior on the optical depth to reionization $\tau$ corresponding to the constraint from Planck PR4 $\tau= 0.051 \pm 0.006$ \cite{planck20-57, SPT-3G:2025bzu}.
     \item \textit{Planck}, specifically the 2018 primary CMB and lensing data based on the \texttt{plik} PR3 likelihood~\cite{planck18-5} and PR4 NPIPE likelihood. 
     \item CMB-SPA: the combination of the SPT-3G D1 SPT\texttt{lite} likelihood \cite{Balkenhol:2024sbv}, ACT DR6 (full multifrequency likelihood), Planck PR4 data with multipole cuts $\ell<1000$ for temperature and $\ell<600$ for polarization, and the Planck \texttt{sroll2} EE likelihood. This combination differs from that studied in \cite{SPT-3G:2025bzu} in several ways:  the use of the full ACT likelihood, the use of Planck PR4 in place of Planck PR3\footnote{Though we note the recent detailed analysis Ref.~\cite{Jense:2025wyg}, which demonstrates the relative insensitivity of cosmological parameter constraints to the choice of Planck likelihood when Planck is combined with other CMB data sets.},  and the use of the \texttt{sroll2} in place of a Gaussian prior on $\tau$. 
 \end{itemize}
In each case we restrict to primary CMB (temperature and polarization) power spectra with CMB lensing used only when explicitly stated; the impact of lensing will be considered in Sec.~\ref{sec:lensing}.

We use BAO data from the DESI DR2 \cite{DESI:2025zgx}. We express the BAO constraints are expressed in the plane of $r_d h$ and $\Omega_m$, and note that this choice is is a lossless compression of the BAO data in the context of $\Lambda$CDM (see \cite{Ferreira:2025lrd} for a discussion).

We perform Markov Chain Monte Carlo analyses of the above data sets using \texttt{Cobaya} \cite{torrado_lewis_2019}\footnote{\url{https://github.com/CobayaSampler/cobaya/tree/master}}, and use \texttt{GetDist} ~\cite{GetDist}\footnote{\url{https://github.com/cmbant/getdist}} to
analyze and plot the results.  We assess convergence of MCMC chains using a Gelman-Rubin statistic and impose a convergence criterion $R-1<0.01$.

We use publicly available MCMC chains when possible, namely for Planck PR3 alone, ACT DR6 alone, and the P-ACT combination of Planck and ACT. We perform independent MCMC analyses in all cases including DESI and in all cases including SPT.

\section{Constraints on $n_s$ from CMB and DESI}
\label{sec:constraints}

Our primary science results are shown in Fig.~\ref{fig:CMB-all-DESI} and Tab.~\ref{tab:constraints-table}, where we show the posterior distributions and parameter constraints from CMB data and from CMB data combined with DESI DR2 BAO. The DESI constraints on $r_dh$ and $\Omega_m$ are shown by grey shaded bands. 

We compare the $n_s$ constraints with the predictions of the Starobinsky, Higgs, and $\alpha$-attractor inflation models, which predict \cite{Kallosh:2025ijd}
\begin{equation}\label{largeN}
    n_s = 1 - \frac{2}{N_*} \ ,
\end{equation}
where $N_*$ is the number of efolds before the end of the inflation that the CMB pivot scale exited the horizon. We take $N_*=[50,60]$, corresponding to a range in $n_s$ of $[0.9600,0.9667]$. This range of $n_s$ is indicated in Fig.~\ref{fig:CMB-all-DESI} by pink bands.

\subsection{Constraints on $n_s$ from CMB data}

\begin{figure*}
    \centering
    \includegraphics[width=0.9\linewidth]{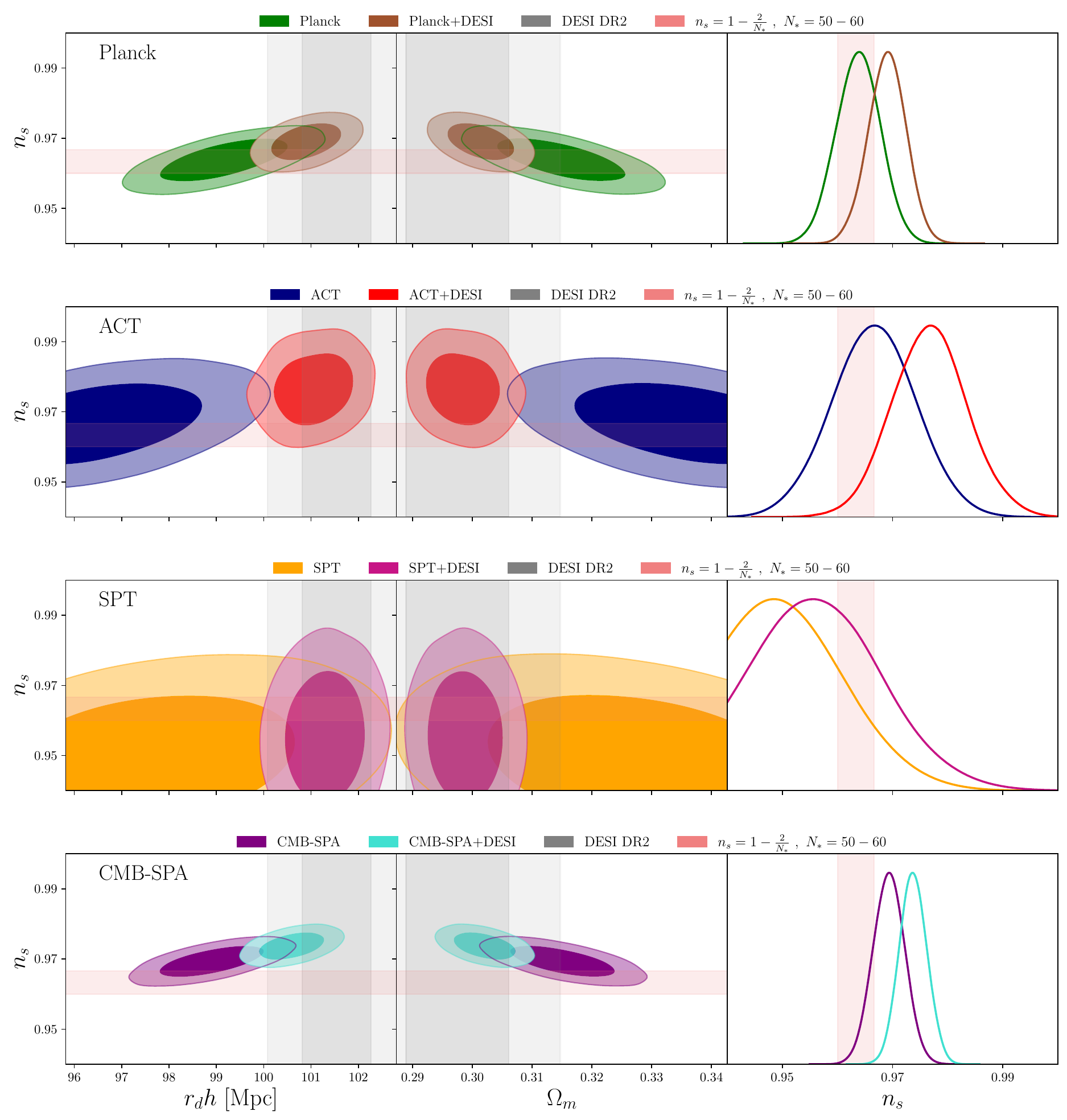}
    \caption{CMB constraints on $n_s$ and the BAO parameters $r_d h$ and $\Omega_m$. We consider CMB data from Planck PR4, ACT, SPT, and the combination CMB-SPA, with and without DESI BAO data. (See Sec.~\ref{sec:datasets} for the details of each dataset). In each case, the inclusion of DESI leads to a shift in $n_s$ due to the correlation of $n_s$ with the BAO parameters in the fit to CMB data. }
    \label{fig:CMB-all-DESI}
\end{figure*}

\begin{table*}
\setlength{\tabcolsep}{12pt}
\centering
\begin{tabular}{@{}cccccc@{}}
\toprule
Data set                  & $n_s$                         & $r_d h$ [Mpc]     & $\Omega_m$      & $\tau$       \\
\toprule 
DESI     & --   & $ 101.54\pm0.73$ & $0.2975\pm 0.0086$  & --\\
\hline
Planck   & $0.9638\pm 0.0040$  & $99.16\pm 0.88$ & $0.3149\pm 0.0070$ & $0.0516\pm 0.0078$  \\
Planck + DESI & $0.9690\pm 0.0035$ & $100.90\pm 0.49$ &  $0.3014\pm 0.0037$ & $0.0554\pm 0.0076$ \\ 
\hline
ACT    & $0.9666\pm 0.0076$   & $96.5\pm 1.5$  & $0.337\pm 0.013$ & $0.0562\,^{+0.0053}_{-0.0063}$ \\
ACT + DESI & $0.9767\pm 0.0068$ &$101.05\pm 0.55 $ & $0.2984\pm 0.0041 $ & $ 0.0636\pm 0.0064$  \\
\hline
SPT   & $0.948 \pm 0.012$ & $97.6 \pm 2.1$ &  $0.328\pm 0.017$ & $0.0506\pm 0.0059$ \\
SPT  + DESI  & $0.955\pm 0.012$ & $101.30\pm 0.56$ & $0.2988\pm 0.0042$ & $0.0529\pm 0.0059$ \\
\hline
CMB-SPA      & $0.9693\pm 0.0029$ & $98.90\pm 0.72$ & $0.3151\pm 0.0057$& $0.0557\pm 0.0036$ \\
CMB-SPA + DESI  & $0.9737\pm 0.0025$ & $100.59\pm 0.45$ & $0.3021\pm 0.0034$ & $0.0575\pm 0.0036$ \\
\botrule
\end{tabular}
\caption{ \label{tab:constraints-table}
Constraints on the scalar spectral index $n_s$, the BAO parameters $r_d h$ and $\Omega_m$ from CMB experiments, and the optical depth $\tau$. The CMB datasets considered are CMB temperature and polarization power spectra from Planck PR4, ACT DR6, SPT-3G DR1, and the ``CMB-SPA'' combination of SPT, Planck ($\ell <1000$ for TT and $\ell<600$ for EE and TE), and ACT, without lensing. See Sec.~\ref{sec:datasets} for the detailed implementation of each dataset.
}
\end{table*}

We first focus on the CMB only results. We stress that ACT, SPT, and Planck data are all in excellent agreement, in the context of $\Lambda\mathrm{CDM}$ and in extensions \cite{SPT-3G:2025bzu, ACT:2025tim, ACT:2025fju}. All three experiments are compatible with the prediction for $n_s$ made by Starobinsky, Higgs, and $\alpha$-attractor inflation models as indicated by pink bands in Fig.~\ref{fig:CMB-all-DESI}.

Planck places the tightest constraint on $n_s$, while SPT places a significantly looser constraint than either ACT or Planck. This can in part be attributed to the limited sky area ($f_{sky}\approx 4\%$) of SPT (see also Fig.~2 of \cite{SPT-3G:2025bzu}). 
The combination of all three, CMB-SPA, largely tracks the Planck constraint, albeit shifted away from DESI, pulled mainly by ACT.

From Fig.~\ref{fig:CMB-all-DESI}, one may appreciate that all CMB experiments, individually and in combination, exhibit a correlation between $n_s$ and the BAO parameters $r_dh$ and $\Omega_m$. As explained in \cite{Ferreira:2025lrd}, these correlations are inherited from the partial degeneracy between $n_s$ and the physical matter density $\omega_m  \equiv \Omega_m h^2$. For example, a constraint on $\omega_m$ can be combined with a constraint on $\theta_s$, the latter being sensitive to the combination $\Omega_m h^3$ \citep{2dFGRSTeam:2002tzq} but independent of $n_s$, to jointly constrain $\Omega_m=\omega_m^3/(\Omega_m h^3)^2$, resulting in an $n_s-\Omega_m$ degeneracy. The physical matter density similarly impacts $r_d$ and $h$, and is inversely correlated with both of these, leading again to correlations with $n_s$.

The degree of correlation depends on the CMB experiment. Paralleling the $n_s$ constraint, SPT exhibits the weakest correlations while Planck exhibits the strongest. These correlations arise from an interplay of the range of angular scales probed by each experiment and the relative emphasis on temperature versus polarization set by the noise levels. From Fig.~\ref{fig:CMB-all-DESI} one may infer that experiments focusing on progressively smaller angular scales tend to exhibit reduced correlations between these parameters.

This trend can be understood from two physical effects: (1) radiation driving of CMB acoustic oscillations \citep{Hu:1996mn,Planck:2016tof}, and (2) gravitational lensing. Radiation driving is sensitive to the physical matter density $\omega_m = \Omega_m h^2$, but the constraining power, and the degeneracy with $n_s$, primarily comes from large angular scales.  In contrast, gravitational lensing, the imprint of gravitational lensing on CMB temperature and polarization anisotropies, is more important (makes a larger relative contribution to the power spectrum) on {\it smaller} angular scales, providing a constraint on $\Omega_m h^2$ that is relatively uncorrelated with $n_s$. The net result of these two effects is a correlation between $n_s$ and the BAO parameters $\Omega_m$ and $r_d h$ that depends on the relative sensitivity to large vs.~small angular scales.

\subsection{Constraints on $n_s$ from CMB+DESI}

\begin{figure*}
    \centering
    \includegraphics[width=0.9\linewidth]{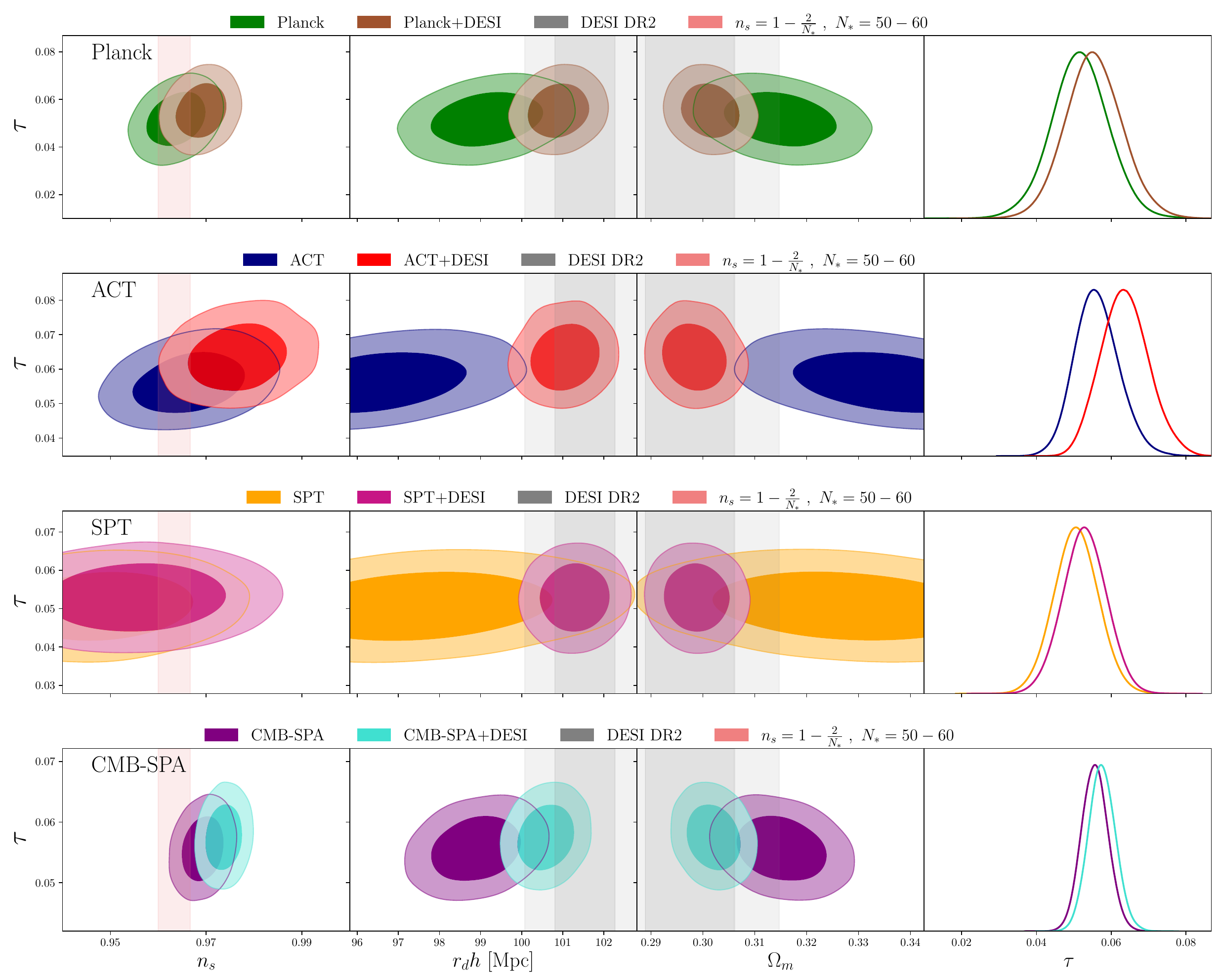}
    \caption{Interplay with $\tau$: CMB constraints on $\tau$, the spectral index $n_s$, and the BAO parameters $r_d h$ and $\Omega_m$. We consider CMB data from Planck, ACT, SPT, and the combination CMB-SPA, with and without DESI BAO data. Paralleling $n_s$, in each case, the inclusion of DESI leads to a shift in $\tau$ due to the correlation of $\tau$ with the BAO parameters in the fit to CMB data.}
    \label{fig:CMB-all-DESI-tau}
\end{figure*}

We now turn to constraints when CMB data is combined  with DESI BAO data, discussing each of Planck, ACT, SPT, and CMB-SPA in turn. We again refer to Fig.~\ref{fig:CMB-all-DESI} and Tab.~\ref{tab:constraints-table}. We note that the combination SPT+DESI, where SPT refers to the full SPT likelihood for primary CMB and without lensing, differs from that reported in \cite{SPT-3G:2025bzu} where SPT+DESI refers to \texttt{SPT-lite}+lensing+DESI. In Sec.~\ref{sec:lensing} we will discuss the impact of lensing data and present the first results from the full SPT likelihood in combination with SPT lensing data and DESI DR2 BAO data.

Given the degeneracy between the BAO parameters and $n_s$ in the fit to CMB, and the tension between CMB and DESI BAO data, one naturally expects that the combined dataset of CMB+DESI will exhibit a shifted $n_s$ constraint relative to that from CMB alone, as the BAO parameters shift from their CMB-preferred value closer to the DESI constraints, pulling $n_s$ along with them.

This expectation is borne out in Fig.~\ref{fig:CMB-all-DESI}, where one may appreciate that for each of Planck, ACT, SPT, and the combination CMB-SPA, the CMB+DESI constraint on $n_s$ is shifted upwards relative to the CMB constraint.  ACT exhibits the largest shift in the central value of $n_s$, but the uncertainty $\sigma(n_s)=0.0068$, is such that the ACT+DESI constraint remains compatible with the Starobinsky, Higgs, and exponential $\alpha$-attractor inflation models.
Comparing with SPT, we note that SPT exhibits a smaller shift in $n_s$, which can be understood from the weaker correlation between $n_s$ and the BAO parameters, however the upward shift from DESI acts to push the SPT $n_s$ constraint {\it towards} the inflation model predictions, since SPT favors a relatively lower value of $n_s$ than ACT. In both the cases of SPT and ACT the inclusion of DESI re-orients the $n_s$-$r_dh$ and $n_s$ - $\Omega_m$ contours, leading to an overall decorrelation of $n_s$ and the BAO parameters in the fit to the combined data set.

Planck exhibits the tightest correlation between $n_s$ and the BAO parameters and the tightest constraint on $n_s$. On the other hand,  Planck exhibits a relatively mild ($2.0\sigma$) discrepancy with DESI. 
When combined with DESI, the Planck $n_s$ shifts upwards, but remains compatible with the Starobinsky, Higgs, and $\alpha$-attractor inflation models. 
The Planck data are constraining enough that $n_s$ and the BAO parameters remain highly correlated even after including DESI, unlike for SPT and ACT.

Finally, looking at CMB-SPA, shown in the bottom row of Fig.~\ref{fig:CMB-all-DESI}, one can see the $n_s$ correlations in CMB-SPA are very well aligned with Planck (which has the most constraining power on $n_s$) and with ACT. When combined with DESI, the CMB-SPA constraints mirror Planck and ACT, with an even smaller error bar on $n_s$ than Planck. As a result, the Starobinsky, Higgs, and exponential $\alpha$-attractor inflation models are disfavored at more than $2\sigma$.

\section{Interplay of $n_s$ and $\tau$}
\label{sec:tau}

\begin{figure}[h!]
    \centering
    \includegraphics[width=\linewidth]{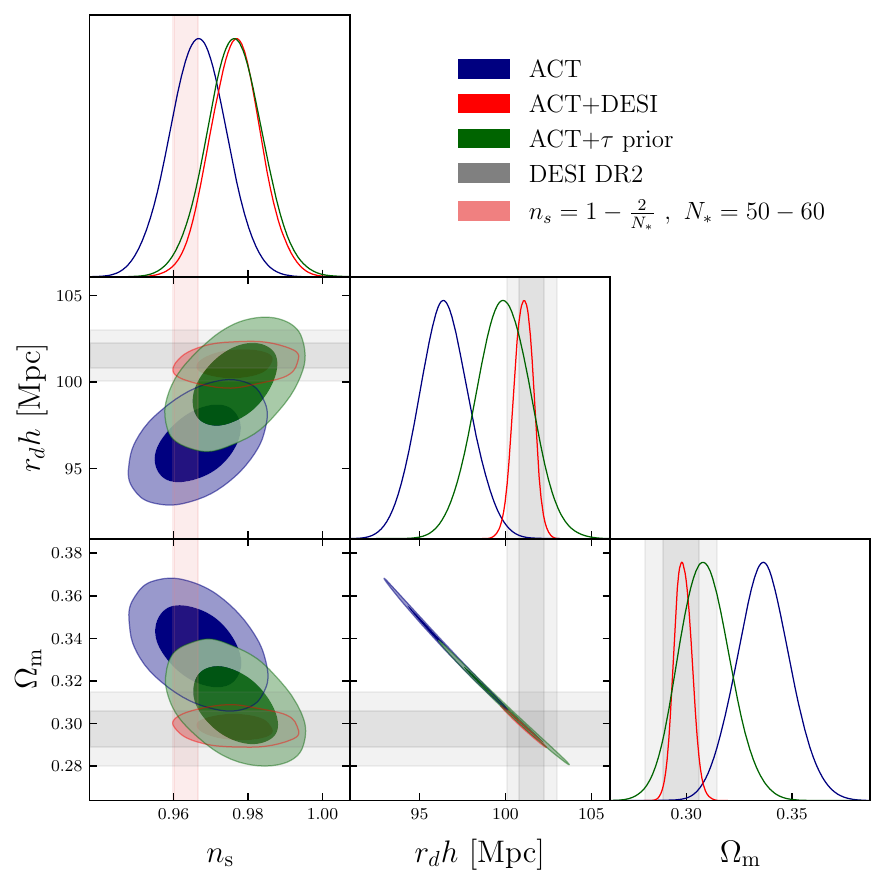}
    \caption{Influence of an artificial $\tau$ prior on ACT. The parameter shifts mirror that from the addition of DESI, and the $n_s$ constraint matches identically.}
    \label{fig:ACTtaupriortriangle}
\end{figure}

The role of $\tau$ in reconciling the tension between CMB and BAO data has been studied in e.g.~\cite{Jhaveri:2025neg,Liu:2025bss,Allali:2025yvp,Sailer:2025lxj}. Here we highlight the correlation with $n_s$.

The spectral index $n_s$ is partially degenerate with the optical depth to reionization, and constraints on these two parameters have shifted in tandem as data sets have evolved (see e.g.~\cite{Planck:2018jri}). Physically this arises because an increase in $\tau$ suppresses power particularly at high-$\ell$, which can be compensated by an increase in $n_s$.  
Additionally $\tau$ is tightly correlated with $A_s$, and $A_s$ is in turn tightly correlated with $n_s$ on large angular scales, adding to the $\tau-n_s$ degeneracy.

Fig.~\ref{fig:CMB-all-DESI-tau} shows the $\tau$ constraint and correlations with $n_s$ and BAO parameters with and without DESI. Constraints on $\tau$ are given in the last column of Tab.~\ref{tab:constraints-table}. From this one may appreciate that the shifts in $n_s$ are accompanied by shifts in $\tau$. The effect is most dramatic for ACT, while Planck and CMB-SPA both exhibit a milder ($\approx0.5\sigma$) shift in $\tau$ in units of the constraint without DESI. We note that  ``ACT'' refers to ACT DR6 primary CMB combined with Planck \texttt{sroll2} low-$\ell$ EE. The differing shifts in $\tau$ for Planck vs.~ACT highlight the important role of Planck beyond low-$\ell$ EE in constraining $\tau$ when Planck is combined with DESI. 

Motivated by this, we perform a simple exercise: we reanalyse ACT with \texttt{sroll2} replaced by an artificial prior  on $\tau$ given by a Gaussian distribution of mean $0.11$ and standard deviation $0.011$.  The posterior distributions are shown in Fig.~\ref{fig:ACTtaupriortriangle}. From this one may appreciate that an artificial upward shift in $\tau$ can match the ACT+DESI $n_s$ and shift BAO parameters to significantly overlap with the ACT+DESI contours. Thus the influence of DESI can in large part be modeled as a prior on $\tau$, despite DESI itself playing no role in constraining $\tau$. 

While we do not claim that this prior on $\tau$ is physical, we note that constraints on $\tau$ are dependent on the model of reionization, and in certain models $\tau$ can be much larger than the CMB constraint, for example, Ref.~\cite{Tan:2025cua,Tan:2025obi} finds $\tau\sim 0.1$ in the population III.1 theory for supermassive black hole (SMBH) formation. Further discussion of the link between the physics of reionization and constraints on $\tau$ can be found in e.g.~Refs.~\cite{Montero-Camacho:2024dzs,Kageura:2026ryq}.

Finally we note that $\tau$ is also correlated with the tensor-to-scalar ratio $r$, due to the correlation of both of these with the scalar amplitude $A_s$. Recent constraints on the $r$ can be found in \cite{Balkenhol:2025wms}.

\section{The Role of CMB Lensing Data}

\label{sec:lensing}

\begin{figure*}
    \centering
    \includegraphics[width=.9\textwidth]{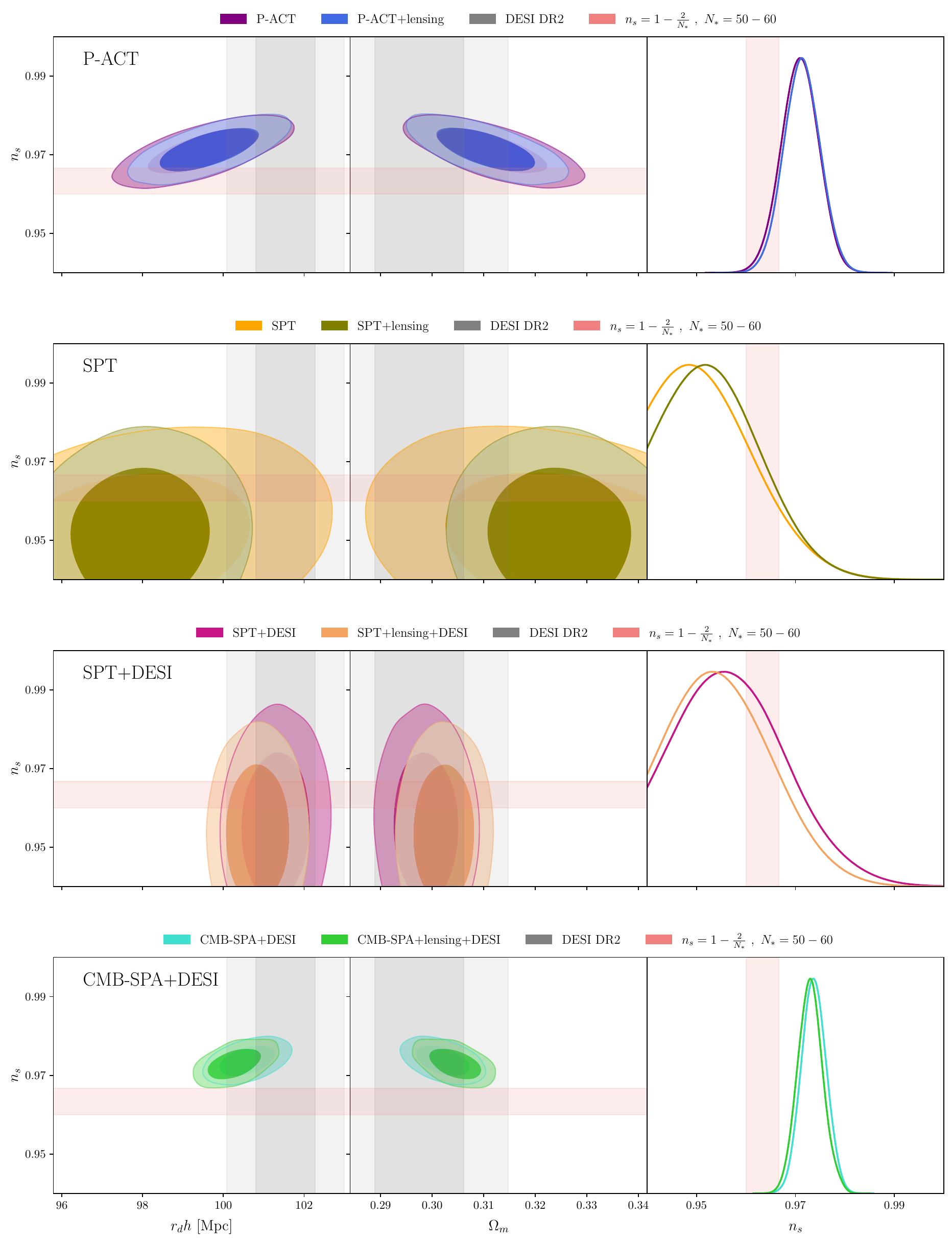}
    \caption{Constraints with and without CMB lensing. We consider P-ACT,  SPT, SPT+DESI, and CMB-SPA+DESI. The inclusion of lensing makes a negligible change to the $n_s$ constraint. In the case of SPT, lensing lessens the correlation between $n_s$ and BAO parameters, but removes most of the overlap with DESI and SPT contours, indicating an increase in the BAO tension. This leads to a slight relative increase in $n_s$ when SPT is combined with DESI with CMB lensing included.
    }
    \label{fig:lensing}
\end{figure*}

\begin{table*}
\setlength{\tabcolsep}{12pt}
\centering
\begin{tabular}{@{}cccccc@{}}
\toprule
Data set                  & $n_s$                         & $r_d h$ [Mpc]     & $\Omega_m$   & $\tau$          \\
\toprule 
P-ACT    & $0.9709\pm 0.0038$   & $99.49\pm 0.91$  & $0.3117\pm 0.0071$ & $0.0604^{+0.0055}_{-0.0065}$ \\
P-ACT + lensing & $0.9713\pm 0.0036$ & $99.64\pm 0.82$ & $0.3105\pm 0.0064$ & $0.0604^{+0.0054}_{-0.0066}$ \\
\hline
SPT   & $0.948 \pm 0.012$ & $97.6 \pm 2.1$ &  $0.328 \pm 0.017$ & $0.0506\pm 0.0059$\\
SPT+lensing & $0.951 \pm 0.011$ & $97.9 \pm 1.1$ & $0.3247 \pm 0.0091$ & $0.0507\pm 0.0059$ \\
\hline
SPT  + DESI  & $0.955\pm 0.012$ & $101.30\pm 0.56$ & $0.2988\pm 0.0042$ & $0.0529\pm 0.0059$ \\
SPT+lensing+DESI  & $0.953 \pm 0.012$ & $100.84 \pm 0.52$ & $0.3023 \pm 0.0039$ & $0.0567\pm 0.0056$\\
\hline
CMB-SPA  + DESI  & $0.9737\pm 0.0025$ & $100.59\pm 0.45$ & $0.3021\pm 0.0034$ & $0.0575\pm 0.0036$ \\
CMB-SPA+lensing + DESI   & $0.9730\pm 0.0026$ & $100.28\pm 0.43$ & $0.3044\pm 0.0033$ & $0.0587\pm 0.0042$ \\
\botrule
\end{tabular}
\caption{ \label{tab:spt_constraints}
Cosmological parameter constraints with and without CMB lensing data. For P-ACT and P-ACT+lensing we use the publicly available MCMC chains. 
}
\end{table*}

The analysis of the previous section focused on the primary CMB power spectra, namely the CMB temperature and polarization anisotropies, but and did not include the power spectrum of the reconstructed lensing potential. The reason for this choice was to allow for a simple physical interpretation in terms of CMB physics, such as radiation driving and gravitational lensing, and to avoid the nuances of lensing reconstruction. Here we investigate quantitatively the role of CMB lensing-potential power spectra in constraining the spectral index $n_s$. Results are shown in Figures \ref{fig:lensing} and Tab.~\ref{tab:spt_constraints}.

In Fig.~\ref{fig:lensing} we compare the constraints on $n_s$, $r_d h$, and $\Omega_m$, from CMB data with and without the CMB lensing likelihood included. Specifically we consider SPT and the ``P-ACT'' combination of ACT and Planck ${\ell}<1000$ for TT and $\ell<600$ for polarization (see Ref.~\cite{ACT:2025fju} for details). We combine P-ACT with ACT DR6 lensing data \cite{ACT:2023kun,ACT:2023dou} and Planck lensing \cite{Carron:2022eyg}\footnote{See \texttt{\url{https://github.com/ACTCollaboration/act_dr6_lenslike}} for details.}, and we combine SPT with the SPT MUSE lensing data \cite{muse}. We also consider the combination CMB-SPA with lensing from Planck, ACT, and SPT.

From Fig.~\ref{fig:lensing} one may appreciate that, for the case of P-ACT, the addition of lensing makes a negligible change to the $n_s$ and BAO parameter constraints, and the change in contours is barely visible by eye. 

In contrast, when SPT CMB data is analysed with lensing included there is a significant reduction in the uncertainty on the BAO parameters (see Tab.~\ref{tab:spt_constraints}), by a factor of $\approx 2$ for both $r_d h$ and $\Omega_m$, which increases the relative tension with DESI. The inclusion of lensing also leads to a relative de-correlation of BAO parameters with $n_s$, and a small but noticeable upward shift in $n_s$.

To study the role of lensing in constraints when DESI is included, we consider constraints from CMB+DESI with and without CMB lensing, for the CMB data sets SPT and CMB-SPA.
The results are shown in the bottom rows of Fig.~\ref{fig:lensing} and Tab.~\ref{tab:spt_constraints}. From this one may appreciate that the inclusion of CMB lensing data does not significantly modify the $n_s$ constraints.

\section{Discussion}
\label{sec:discussion}

In this work, we have endeavored to understand the spectrum of $n_s$ constraints arising from CMB data in combination with BAO data from DESI DR2. The science results of this work can be summarized as follows: In the context of the $\Lambda$CDM model,
\begin{itemize}
    \item All CMB experiments exhibit a correlation between $n_s$ and BAO parameters $r_d h$ and $\Omega_m$. The strength of the correlation depends on the relative sensitivity to large vs.~small angular scales of each experiment.
    \item All CMB experiments exhibit a shift in $n_s$ when CMB data is combined with DESI BAO. The largest relative shift in $n_s$ occurs for ACT, driven by the tension with DESI which is the largest of the CMB experiments considered. When the CMB experiments are combined into CMB-SPA, and together combined with DESI, the constraint on $n_s$ disfavors the Starobinsky, Higgs, and exponential $\alpha$-attractor inflation models at $>2\sigma$. 
    \item The shifts in $n_s$ are accompanied by shifts in $\tau$. The shift in $n_s$ when ACT is combined with DESI can be mimicked by imposing an artificial prior $\tau \sim 0.1$, which is much larger than the Planck constraint. 
    \item The inclusion of the CMB lensing data does not significantly change the $n_s$ constraints from CMB+DESI. 
\end{itemize}
A natural question is the model dependence of these results. Indeed several models are able to ameliorate the BAO-CMB tension \cite{SPT-3G:2025bzu}, and each of these are favored over $\Lambda$CDM at the $2-3\sigma$ level. The DESI data has been intensely studied as possible evidence for dynamical dark energy, and in particular the combination of Planck and DESI DR2, in the $w_0w_a$ model with broad uniform priors on $w_0$ and $w_a$, gives a preference for dynamical dark energy at greater than  $3\sigma$. For a discussion of prior dependence of the DESI evidence for dark energy see e.g.~\cite{Toomey:2025xyo,Payeur:2024dnq, Wolf:2024eph, Ramadan:2024kmn, Akrami:2025zlb}. The Planck+DESI constraint on $n_s$ in the $w_0w_a$ model, for various choices of priors, is given in Tab.~II of Ref.~\cite{Toomey:2025xyo} and is given by $n_s=0.9638 \pm 0.0038$ for broad uniform priors on $w_0$ and $w_a$, nearly identical from the $\Lambda$CDM constraint from Planck alone, and 
in excellent agreement with the Starobinsky, Higgs, and exponential $\alpha$-attractor inflation models.
This provides an indication that the CMB+DESI constraint on $n_s$ is entangled with the DESI evidence for dynamical dark energy.

The model dependence of $n_s$ constraints can be compared with that in resolutions to the the Hubble tension: When Planck data is combined with SDSS BAO, Pantheon supernovae, redshift space distortions, and $H_0$ from SH0ES (circa 2020), the constraint in $\Lambda$CDM is $n_s=0.9689 \pm 0.0036$, consistent with the result from Planck alone, while in the Early Dark Energy proposal to resolve the Hubble tension, the constraint on $n_s$ shifts upward, to  $n_s=0.9854 ^{+0.0070} _{-0.0069}$ for the same data sets \cite{Hill:2020osr,McDonough:2023qcu}. (For an analysis of EDE with ACT DR6, see \cite{AtacamaCosmologyTelescope:2025nti}, and for CMB-SPA, see Ref.~\cite{SPT-3G:2025vyw}).
The situation with Early Dark Energy and the Hubble tension is therefore the reverse as that presented here: it is the extension to $\Lambda$CDM which drives the increase in $n_s$.

We reiterate the importance of $n_s$ for the status of inflation \cite{Kallosh:2025ijd,Linde:2025pvj,Ferreira:2025lrd}: the spectral index $n_s$ has been and continues to be a powerful test of inflation models. We again emphasize that {\it many} well motivated inflation models predict a tensor-to-scalar ratio that falls below the threshold for next generation CMB experiments, and would be more directly tested by more precise measurements of $n_s$ and the running $\alpha_s$  \cite{Lorenzoni:2024krn,Hardwick:2018zry,Martin:2024nlo}. This fact, and the results of this paper, motivate the continued experimental effort to constrain $n_s$. Looking to the future, large scale structure experiments such as Euclid \cite{Euclid:2025dlg} will dramatically improve the precision of BAO measurements and therefore may play an important role in determining the fate of $n_s$. Future work is also needed to include existing datasets, such as Dark Energy Survey, into the analysis presented here, as well as current and upcoming Lyman-$\alpha$ data

Finally, we note the reanalysis of DESI data performed in \cite{Chudaykin:2025lww,Chudaykin:2025aux} may have important implications for constraints on $n_s$.

We leave this and other interesting directions to future work.

\acknowledgments

The authors thank Lennart Balkenhol, J.~Colin Hill, Mikhail Ivanov, Renata Kallosh, LLoyd Knox, Eiichiro Komatsu, Andrei Linde, Paulo Montero-Camacho, and Leander Thiele for helpful discussions. E.M. is supported in part by a Discovery Grant from the Natural Sciences and Engineering Research Council of Canada, and by a New Investigator Operating Grant from Research Manitoba. Kavli IPMU is supported by the World Premier International Research Center Initiative (WPI), MEXT, Japan. EGMF thanks the support of the Serrapilheira Institute. \\

\appendix

\section{Comparison of Planck PR3 and PR4 NPIPE}

We repeat the analysis of Planck and Planck+DESI  for the Planck PR3 data using the \texttt{plik} likelihood. Parameter constraints are given in Tab.~\ref{tab:constraints-table-PR3-PR4}.  

\begin{table*}
\setlength{\tabcolsep}{12pt}
\centering
\begin{tabular}{@{}cccccc@{}}
\toprule
Data set                  & $n_s$                         & $r_d h$ [Mpc]     & $\Omega_m$      & $\tau$       \\
\toprule 
Planck PR3   & $0.9649\pm 0.0044$  & $98.9\pm 1.0$ & $0.3166\pm 0.0084$   & $0.0544^{+0.0070}_{-0.0081}$ \\
Planck PR3 + DESI & $0.9713\pm 0.0033$ & $101.02\pm 0.51$ &  $0.3003\pm 0.0038$ & $0.0582\pm 0.0080$ \\ 
\hline
Planck  PR4  & $0.9638\pm 0.0040$  & $99.16\pm 0.88$ & $0.3149\pm 0.0070$ & $0.0516\pm 0.0078$  \\
Planck PR4 + DESI & $0.9690\pm 0.0035$ & $100.90\pm 0.49$ &  $0.3014\pm 0.0037$ & $0.0554\pm 0.0076 $\\ 
\botrule
\end{tabular}
\caption{ \label{tab:constraints-table-PR3-PR4}
Comparison of Planck PR3 and PR4 with and without DESI.}
\end{table*}

\bibliography{refs}

\end{document}